**Heart Valve Collaboratory – Heart Valve Component Fatigue Demonstration Testing**

**Attribute and Variables Demonstration Test Planning and Sample Size Determination**

Wayne M Falk[1] and William Q. Meeker[2]

1.  Abstract

The fatigue demonstration test program may be carried out using either the attribute or variables approach.  For either approach, a robust test program fulfills the objective of unambiguously demonstrating reliability, as well as, demonstrating understanding of when and how the components could fracture.  To this end, it is recommended that a robust test program should aim to produce data for both surviving units and fractured units.

Test planning is needed to determine sample sizes and test levels of strain to prove that the component meets the target reliability.  The two risks that must be addressed by the test plan are 1) the risk that an unreliable component will pass the test, i.e., Type 1 error, or 2) the risk of a test failing for a sufficiently reliable component, i.e., Type 2 error.

Test planning methods are described to design a test plan that efficiently meets these objectives for attribute and variable approaches.
- The attribute approach offers the advantage of straight forward planning and a simple output. Sample size is calculated using a well-known formula.  Testing to fracture is recommended to complement the traditional attribute testing approach to identify the test failure modes.
- The variables approach is more complex to plan and analyze, however it enables a greater understanding of reliability.  Sample sizes and test levels of strain are determined using a simulated data strategy to provide a clear picture of possible outcomes of the test.

**Keywords:** Fatigue, Demonstration, Planning, Heart Valves, Reliability, Statistics

2.  Introduction

The fatigue demonstration test program is intended to show that the structural components of a new heart valve will reliably survive in vivo loading conditions for its intended lifetime.

There are two primary outputs of demonstration testing. First, the manufacturer must demonstrate an understanding of the fatigue failure modes of the component.  That knowledge is employed to justify clinically relevant and validated test methods. Second, the manufacturer must generate fatigue data to establish that structural components meet reliability acceptance criteria when subjected to a sufficiently challenging test program.

---

1  U.S. Food and Drug Administration (FDA), Silver Spring, MD, USA. wayne.falk@fda.hhs.gov

2 Iowa State University, Ames, IA USA. wqmeeker@iastate.edu

A test program includes a series of tests conducted to recreate or exceed the in vivo stress or strain magnitude in a clinically relevant loading mode.  The demonstration test requires careful development and documentation of test design, procedures, analysis, feasibility testing, and acceptance criteria.  Demonstration testing is generally best performed on whole devices or whole structural components.  Extracted high-stress/strain portions of finished components may also be used when it is not feasible to reproduce the clinical loading on the entire component or device.

3.   Demonstration Test Approaches

Demonstration test approaches fall into two categories depending on whether the data are attribute or variables data.  In attribute demonstration testing, the outcome of every unit is binary – either a fracture or a survival.  In variables demonstration testing, the outcome of each test is the measured life, i.e., the number of cycles at which fracture occurred at a given test level of strain.  If the unit survives to the end of the test, then the life of that data point is right censored at the test duration (i.e., the life is only known to be greater than the test duration).

The attribute approach to demonstration testing is the most common because of the simplicity of the test plans and the straightforward interpretation of the results. However, the variables approach has advantages.  Variables demonstration tests give more resolution of the reliability of the component throughout the loading range and lifetime. This greater resolution comes at a cost in terms of the number of units to be tested and the greater complexity, particularly in the statistical methods used in both test planning and data analysis.

4.   Essential Elements of Any Test Program

Each fatigue test unit may result in generating fracture data or survival data. Both data types are essential to the understanding of the component's fatigue performance.   Therefore, a robust test program should aim to produce data for both surviving units and fractured units. Although there is no single demonstration test program that will be appropriate for all components, it is recommended that the test program should result in having both fracture and survival units.

Generating fracture data is an essential part of any demonstration test program to address the following questions: Is this a clinically relevant test? What is the component failure mode? How does the observed experimental failure mode compare to the predicted failure mode from the stress analysis?[1] Even in the case of attribute demonstration testing, it does not suffice to only produce units that survive the test.  It recommended that additional tests be run, with levels raised sufficiently above physiological conditions to drive the units to fracture.  The objective of this is to demonstrate when and how the components would fracture, thus giving an assurance that the test is recreating a clinically relevant loading mode.

Conversely, survival data is an essential output, even in the case of a variables test program. The test owner needs a sufficient number of units to survive a challenging strain level to the design life, i.e., without extrapolation, to unambiguously demonstrate reliability.

The proportion of fractures or survivals in tested units depends on the design of the test program and whether the demonstration is an attribute demonstration test or a variables demonstration test.

- In an attribute demonstration test program, a few units are fractured to demonstrate the clinical relevance of the test and an understanding of the failure modes.  A larger, statistically chosen number of units survive to the design life and are used to meet reliability acceptance criteria.
- In a variables demonstration test program, a statistically chosen number of units are tested at multiple strain levels.  Most of the units are intended to fracture.  In addition, a predetermined number should survive to the design life at strain levels seen in use conditions. These data are used to estimate the relationship between lifetime and stress/strain for the component. Both types of data can be used to meet reliability acceptance criteria.

In summary, there exists a spectrum of approaches that balance fracture and survival data, but all approaches should include both types of data.

5. Test Planning

Based on the component's risk analysis, a minimum target reliability, $R$, is set for the demonstration test.  Test acceptance criteria are met if the data provides evidence, beyond a reasonable doubt (e.g., a 95% confidence level), that the component exceeds the target reliability.

Test planning should be performed to ensure that sample sizes and test levels of strain are appropriate to demonstrate that the component meets the target reliability.  The two risks that must be addressed by the test plan are 1) the risk that an unreliable component will pass the test, i.e., Type 1 error, or 2) the risk of a test failing for a sufficiently reliable component, i.e., Type 2 error.

Controlling Type 1 error is the foremost priority in test planning, as it has implications for patient safety.  It is standard practice to use a one-sided lower confidence bound on reliability to control Type 1 error risk.  The confidence level, $C = 100(1 - \alpha)\%$, is set based on the component's risk analysis.  This requires that, given all possible outcomes from the experiment, there is at most a probability $\alpha$ that the lower confidence bound for $R$ will be less than the actual reliability.  Therefore, the risk of Type 1 error is less than or equal to $\alpha$. Often, $\alpha$ is chosen to be 0.05, corresponding to 95% confidence. The construction of one-sided lower confidence bounds for both attribute and variables test plans will be described in later sections.

The probability of successful demonstration, i.e., controlling Type 2 error, depends primarily on the extent to which the actual reliability exceeds the target reliability. More specifically, it depends on estimating the probability of any fatigue unit surviving for the design life at the demonstration test level of strain. Unlike the Type 1 error, the required probability of a successful demonstration is controlled at the discretion of the test owner, based on how much risk of test failure is tolerable. For the following discussions, the probability of a successful demonstration, is denoted $100(1 - \beta)\%$, where the risk of Type 2 error is $\beta$. Methods for computing the probability of a successful demonstration for both attribute and variables approaches will be discussed later in this paper.

6.  Attribute Demonstration Test Program Planning

Component fatigue demonstration testing is typically accomplished via attribute testing. Test levels of strain for an attribute test should be chosen to be at or above the use conditions using the methods discussed in [2]. Test-to-success sample sizes will depend upon the target reliability and confidence levels required by a manufacturer's risk assessment procedures. Besides confidence and reliability, the only remaining parameter that enters the sample size calculation is the number of allowable fractures, $f$. Often the test owner will choose to minimize the number of units needed by not allowing any test fractures (i.e., choosing $f$=0 and which is known as a zero-failure demonstration test). In that case, the minimum reliability demonstrated by the test [3] is given by $R = \alpha^{\frac{1}{n}}$ for sample size, $n$, and confidence level, C= $100(1 - \alpha)\%$, therefore the minimum sample sizes is given by the formula

$$n = \frac{\ln \alpha}{\ln R}$$

where the non-integer result is rounded up. In this case all $n$ units must run the entire duration of the test without fracture. It is most common that component fatigue tests require 95C/90R, requiring an attribute test requiring $n$=29 units.

Alternatively, there are advantages to plans which allow some fractures to occur and still meet the acceptance criteria. To determine the sample size, $n$, with a predetermined number of fractures allowed, $f$, we can use the following binomial distribution formula [4]:

$$\alpha = \sum_{i=0}^{f} \binom{n}{i} (1 - R)^i R^{n-i}$$

based upon confidence level, C= $100(1 - \alpha)\%$, and reliability value, $R$. If we were to allow a single fracture, $f$=1, for 95%C/90%R, we would find the required sample size to be 46 units. Therefore, to be 95% confident that the component is 90% reliable, 46 parts must be tested with no more than one fracture. If the same calculation was carried out for $f$=2 and $f$=3 plans we would find that 61 and 75 units would be required, respectively.  The *f*>0 plans have an increased number of units, but also offer greater probability of successful demonstration, as will be shown below. Further discussion is available in [3, 5].

**Table 1 – Samples sizes, n, for attribute test with a given number of allowable fractures, f**

| f | C/R | n |
|---|-----|---|
| 0 | 95%C/90%R | 29 |
| 1 | 95%C/90%R | 46 |
| 2 | 95%C/90%R | 61 |
| 3 | 95%C/90%R | 75 |

Computing the probability of a successful attribute demonstration, $100(1 - \beta)\%$, requires specification of the actual reliability at the test conditions. $R_{\text{actual}}$, and the formula

$$(1 - \beta) = \sum_{i=0}^{f} \binom{n}{i} (1 - R_{\text{actutal}})^i (R_{\text{actutal}})^{n-i}$$

For example, if the actual reliability at the test conditions is $R_{\text{actual}}$= 0.99, then the probability of successfully demonstrating 95C/90R with an *f*=0 (n=29) test plan is 0.747, and with a *f*=1 (*n*=46) test plan is 0.922 This simple calculation shows the additional robustness for *f*>0 test plans in comparison to *f*=0 plans.

In addition to the attribute results taken at the demonstration test level of strain, it is recommended to present fracture data from the demonstration test method at increased strain amplitudes to identify when and where fractures occur. The location of fracture should be compared to predictions based on FEA modeling to determine whether the outcomes of the fatigue tests match FEA predictions.  Units should be tested to fracture to capture the failure modes at important points of the failure envelope.

7.   Variables Demonstration Test Program Planning

Variables demonstration tests offer the ability to quantify fatigue reliability throughout a range of use condition strain amplitude. Variables test plans should be designed to fracture test units at multiple strain amplitude test levels. The testing is carried out on components of fully finished components with a simulated implantation history.  A discussion of this approach is contained in [6].

Planning a variables life demonstration test requires choosing a set of test levels and allocating a number of units to each level. In this section, we will step through a planning procedure and describe the data and analysis tools needed to plan a robust test.

At a minimum, there will be two objectives of the variables demonstration test plan.  First, the test should demonstrate that the design exceeds the minimum reliability at specified confidence requirements.  Second, a minimum number of units should survive use conditions to runout at the design life.  See Table 2 for an example of the planning objectives

**Table 2 – Example of a Planning Objectives**

| Requirement | Objective |
|---|---|
| 95% Confidence and 90% Reliability Fatigue Strength | Minimum fatigue strength must exceed the limiting use condition at 600M cycles with 90% Reliability and 95% Confidence |
| Minimum Number of Survivals to 600M Cycles | A minimum number of units must survive to runout at 600M cycles. The lowest test level of strain must be greater than or equal to limiting use condition. |

Step 1. Seed Data
Baseline information regarding fatigue behavior is needed to plan the test. Therefore, a first step is to obtain *seed data*, in the form of ε-N data from a material test [7] or initial data on the component being tested.  If such data is not available, a pilot test should be run to generate seed data prior to planning the demonstration test. The dataset does not need to meet any specific requirements, although the size and quality of the data may affect the robustness of the test plan.

Step 2. Choosing an Appropriate Statistical Model
Variables test planning is based on fitting a *statistical model* to the seed data and sampling from that distribution.  The statistical model consists of two main elements, a Strain/Life Relationship and a Probability Model.

*Strain/Life Relationship*—The relationship between strain and life follows a curve. Equations such as an inverse power-law or the Coffin-Manson equation are commonly used to specify the strain/life model. Other relationships may also be appropriate [8].

*Probability Model* – A statistical distribution (e.g., a lognormal or Weibull distribution) is needed to describe scatter in fatigue data.  The probability model may be specified as a life distribution or strength distribution. If the fatigue-life model is specified, it induces a fatigue-strength model and vice-versa.

For further information on model selection see [8].

Step 3.
Bayesian or non-Bayesian (e.g., maximum likelihood) methods may be used to fit the model to the seed data [8]. The model fit will provide a statistical distribution which will, in turn, be used to simulate the outcome of proposed test plans.  The use of simulation in test planning is described in detail in [3, 10].

Before proceeding to propose a test plan, it is recommended to assess the model fit to seed data. The results of the model fit can be used to plot the median life curve (Figure 1). This curve should follow the seed data[2] average life at each strain level.

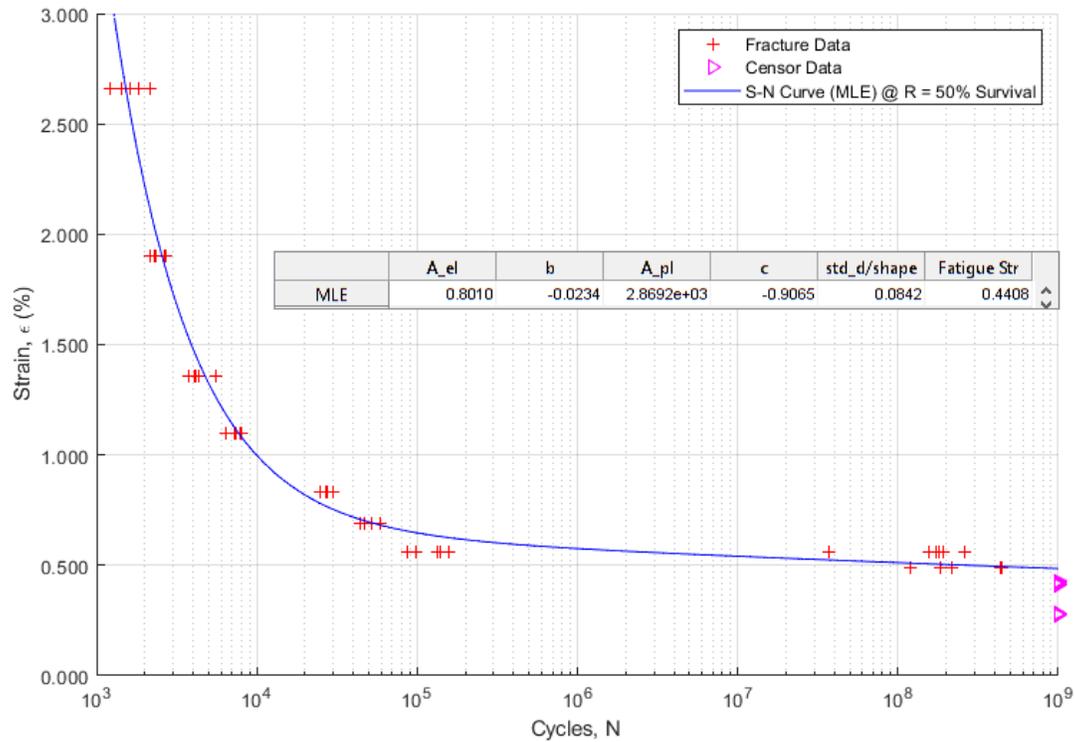

**Figure 1 – Median (50%) life curve**

The statistical model fit will also provide estimates of the reliability life curve of interest for the demonstration test. For example, the 0.10 distribution quantile corresponding to *R*=90% reliability and the corresponding curve is shown in Figure 2. It can be seen graphically that approximately 90% of the data points lie to the right of this 90% reliability curve. In addition to these basic checks, see [8] for additional graphical and quantitative techniques to assess the "goodness of fit" of the statistical model.

---

[2] For illustration, a subset of the data from [9] is used as seed data.

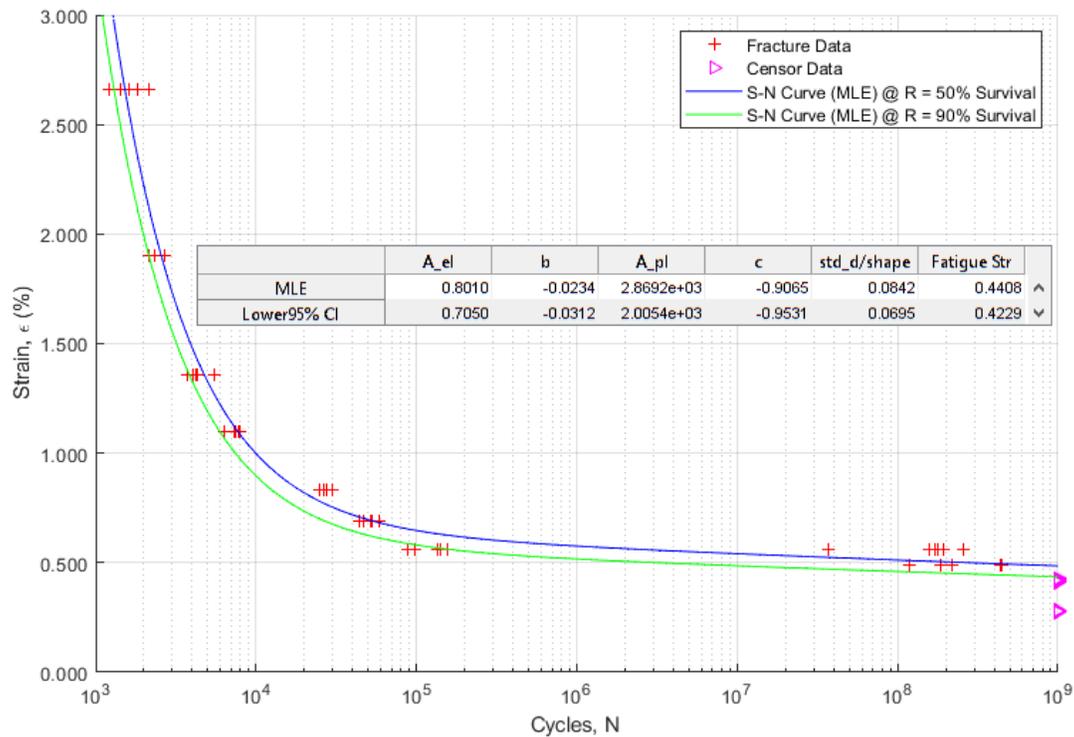

| | A_el | b | A_pl | c | std_d/shape | Fatigue Str |
|---|---|---|---|---|---|---|
| MLE | 0.8010 | -0.0234 | 2.8692e+03 | -0.9065 | 0.0842 | 0.4408 |
| Lower95% CI | 0.7050 | -0.0312 | 2.0054e+03 | -0.9531 | 0.0695 | 0.4229 |

**Figure 2 - Median (50%) and 90% Reliability Curves**

Step 4. Propose Candidate Test Plans

Each candidate test plan consists of a specified set of strain levels, the number of units at each strain level, and the test duration. In general, the candidate plans should meet the following guidelines, unless a rationale is provided to justify an alternative

- Test plans should have at least four test levels of strain.

- The lowest test level must be set at or above the largest use condition, with all units expected to run out to the design life.

- At least the two highest test levels should be targeted to generate fractures to define the failure curve. The highest level should be limited so as to result in the same failure mode(s) as observed at lower levels.

- At least one test level should be chosen to generate at least a few high cycle fractures. This gives the plan the ability to resolve the higher variation present at lower test levels.

- Test plans may be designed to allocate the same number of units to each test level or to allocate more samples to lower, more realistic, test levels. The latter option will more efficiently estimate the device reliability because the probability of failure is smaller at the low levels of strain (information obtained is primarily from failures) and fractures will occur closer to the design life. See Chapter 20 of [10] and references therein.

Multiple test plans may be proposed and compared (e.g., by using Monte Carlo simulation). A plan should be chosen that meets the minimum reliability and confidence objectives of the demonstration test with a high probability of success and with an efficient allocation of units and/or other resources.

For example, consider the following two test plans, shown in Table 3, which are intended to demonstrate that a fictitious component will survive limiting use condition strains of $\varepsilon_{use}$=0.25%, for 600M cycles with 95%C/90%R.

- In Candidate Test Plan 1, 10 units each are tested at 1.00%, 0.75%, 0.50% and 0.25% strain.
- In Candidate Test Plan 2, 5 units each are tested to 0.80% and 0.60% strain and 10 samples each are tested to 0.40% and 0.25% strain.

Test plan 2 utilizes a total of 30 units whereas test plan 1 utilizes 40 units. Both plans will test all units to the 600M cycle test duration. In the following section, we will compare the performance of these plans.

**Table 3 – Example Candidate Test Plans**

Candidate Test Plan 1

| Test Level of Strain | Quantity | Duration |
|---|---|---|
| 1.00% | 10 | 600M |
| 0.75% | 10 | 600M |
| 0.50% | 10 | 600M |
| 0.25% | 10 | 600M |

Candidate Test Plan 2

| Test Level of Strain | Quantity | Duration |
|---|---|---|
| 0.80% | 5 | 600M |
| 0.60% | 5 | 600M |
| 0.40% | 10 | 600M |
| 0.25% | 10 | 600M |

<u>Step 5. Simulate Candidate Test Plans</u>

Candidate test plans are simulated by sampling from the statistical distribution fit to the seed data. For each simulated data point, the sampling algorithm takes as an input a test strain level. A random number is generated from 0 to 1 and mapped to the sample's life (i.e. fracture cycle count) using the inverse CDF of the life distribution at that strain level. If the simulated life is greater than the test duration, then the unit is a runout. Each of the test plans shown in Table 3 was simulated and the resulting fitted models are shown in Figure 3.

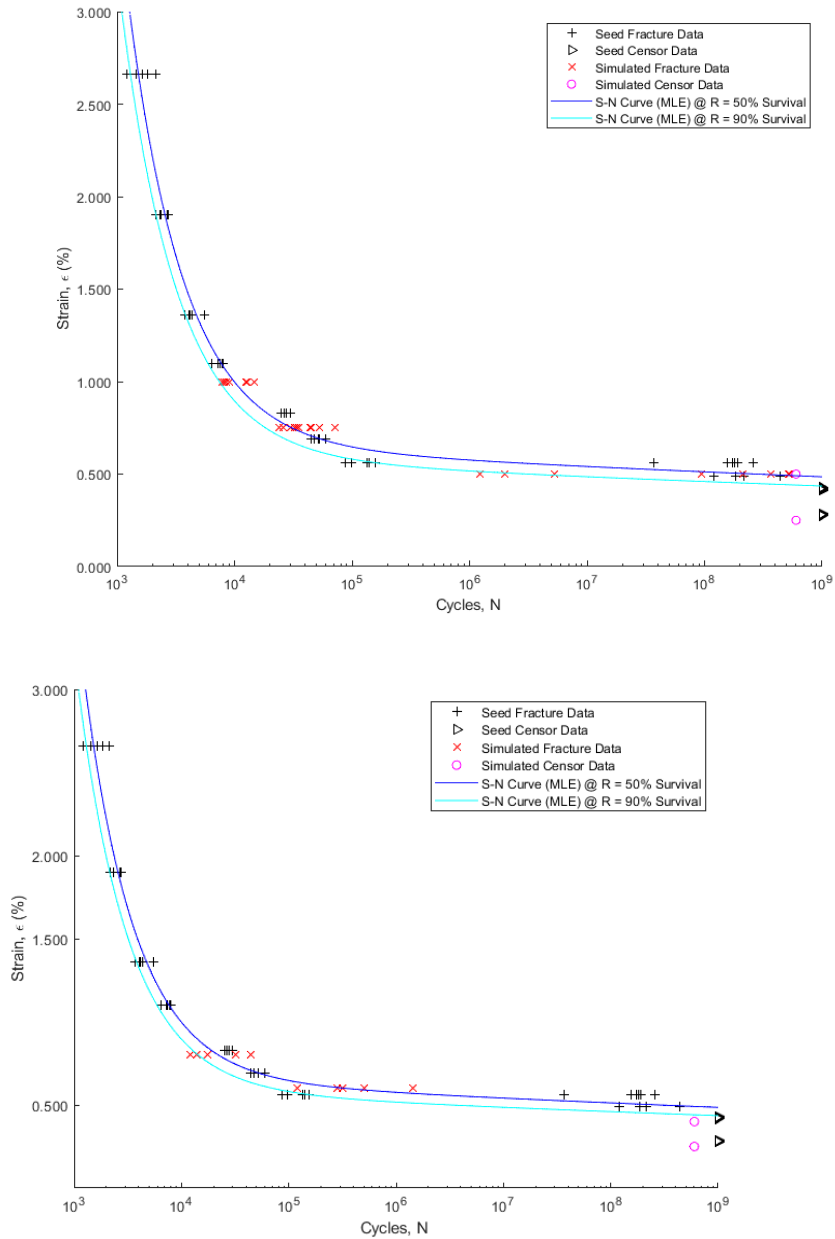

**Figure 3 - Sampling of test plans shown in Table 3 (a) Candidate Test Plan 1 (b) Candidate Test Plan 2.  The original seed data is shown in black and the simulated test results are shown in color.**

Next, the sampled data from candidate test plans are treated as if they were the outcome of variables demonstration tests.  The analysis of the sampled data is used to critically assess each candidate test plan.  Scrutinizing the ability of the data to fully define the fatigue curve may highlight a deficiency of the proposed plan and the need for a larger amount of data in certain strain ranges.  Figure 4 shows the sampled data that was discussed above, analyzed with the statistical model.  As before, the model fit to the sampled data can be assessed using the estimated median life and 90% reliability curve.

Any statistical inference based on a limited amount of data contains statistical uncertainty. For that reason, one-sided lower confidence bounds should be computed to find the most pessimistic model fit to the $R$ quantile that is consistent with the data.  For variables demonstration, as with attribute demonstration, the confidence level C= $100(1 - \alpha)$%, is set based on the component's risk analysis and determines the allowed risk of Type 1 error $\alpha$. Figure 4 shows the lower 95% confidence bound to describe the statistical uncertainty in the estimated R=90% reliability curve fit to the sampled data.  The *lower fatigue strength, $\varepsilon_{strength}$* is taken as the value of the 95% Confidence lower bound on the $R$=90% lower bound line at 600M cycles.  The lower fatigue strength value for a single simulation of each candidate test plan is shown in Figure 4.

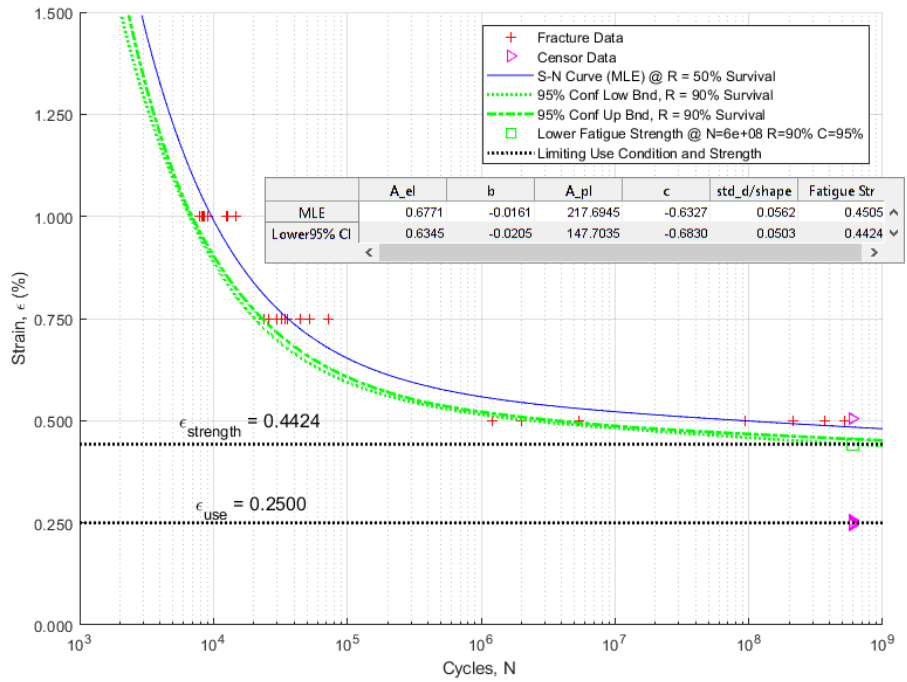

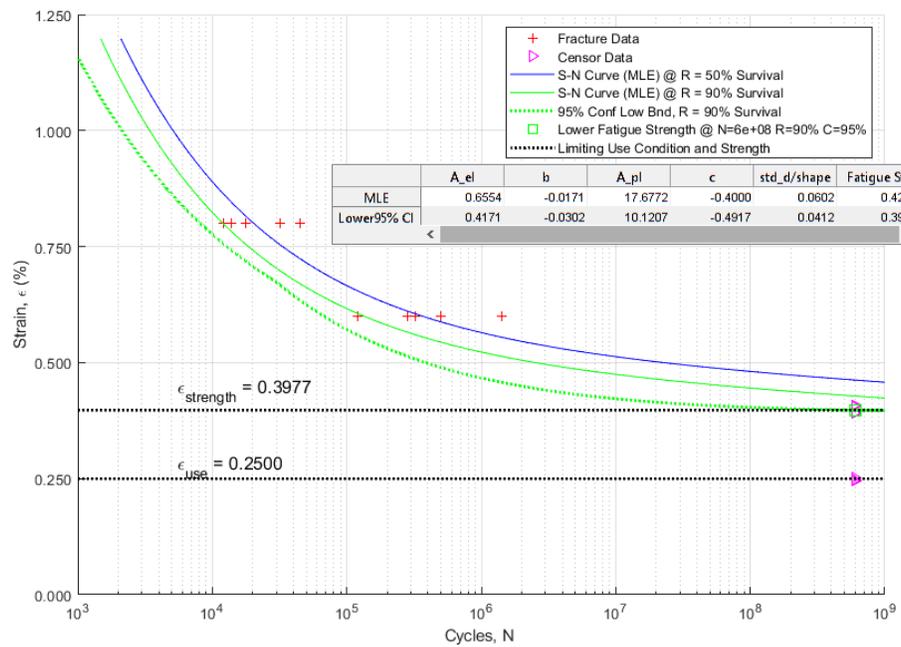

**Figure 4 - Analysis of test plans shown in Table 3 (a) Candidate Test Plan 1 (b) Candidate Test Plan 2**

Step 6. Assess the Performance of the Candidate test plans

Candidate test plans should be simulated repeatedly and compared to explore the range of possible outcomes. Candidate test plans 1 and 2 were each simulated m=5 times to gage the probability of successful demonstration, although a greater number of simulations may be advisable depending on the stability that is observed. The simulation results are shown in Table 4.

**Table 4 – Candidate Test Plan Results**

| | Candidate Test Plan 1 | | Candidate Test Plan 2 | |
|---|---|---|---|---|
| | Confidence and Reliability Fatigue Strength | Number of Survivals to 600M cycles at $\varepsilon_{use}$=0.25% | Confidence and Reliability Fatigue Strength | Number of Survivals to 600M cycles at $\varepsilon_{use}$=0.25% |
| Minimum Target | 0.25% | 10 | 0.25% | 10 |
| Trial 1 | 0.42% | 10 | 0.42% | 10 |
| Trial 2 | 0.42% | 10 | 0.39% | 10 |
| Trial 3 | 0.41% | 10 | 0.50% | 10 |
| Trial 4 | 0.42% | 10 | 0.39% | 10 |
| Trial 5 | 0.43% | 10 | 0.40% | 10 |
| Summary | ALL PASS | ALL PASS | ALL PASS | ALL PASS |

All $m$=5 trials for each of these candidate test plans achieved the objectives of demonstrating a lower bound fatigue strength greater than 0.25% strain and having all 10 units tested at 0.25% strain survive the full 600M cycle test duration. Although there is little difference in the outcomes, test plan 1 has a smaller amount of statistical uncertainty in the fatigue strength quantile estimate. And while test plan 2 requires fewer samples, it did not produce as many high cycle fractures at test plan 1, as shown in the top plot of Figure 3. For these reasons, a slight preference may be given to test plan 1.

The probability of successful demonstration 100(1 – β)%, can be estimated by the following technique. If $m$ simulated trials are performed with $m_s$ of these trials successfully passing, then we can compute (under the assumption that the assumed model is correct) the probability of successful demonstration for variables test plans (expressed as a percent) by

$$100(1 - \beta)\% \gtrsim \frac{100(m_s - 1)}{m}\%$$

Therefore, for both examples above, we would estimate that the probability of success to be greater than 80%.

Step 7. Confirmation of the Candidate test plans

The test owner should be aware that variables test planning depended greatly on the seed data.  Before executing the demonstration test, it is advised to verify that the demonstration test output follows the test simulations. It may be that there is enough discrepancy between the seed data and demonstration test output to cause an unpredicted test outcome.  These differences between the seed data and the demonstration test output could arise due to different material history, strain volume, test variability etc.  For that reason, it is an important step before starting any run of the demonstration test method at the planned test levels of strain, to make sure that the test outputs are falling within the predicted ranges of life at that test level.

8. Summary
   The fatigue demonstration test program may be carried out using either the attribute or variables approach.  In either case, test planning methods are available to select a test plan that efficiently and robustly meets the objectives of the demonstration test program.

9. Conflict of Interest
   The authors declare that there are no conflicts of interest

10. Ethical Approval
    This paper does not contain any studies with human participants or animals.